\documentclass[ascmo, manuscript]{copernicus}

\begin{document}

\title{Predicting Dry Spells of the West African Monsoon Season Using Machine Learning Methods}


\Author[1][cbobo@bu.edu]{Colin}{Bobocea} 
\Author[1][atchade@bu.edu]{Yves}{Atchade}

\affil[1]{Boston University Department of Mathematics \& Statistics}




\runningtitle{TEXT}

\runningauthor{TEXT}

\received{}
\pubdiscuss{} 
\revised{}
\accepted{}
\published{}


\firstpage{1}

\maketitle

\begin{abstract}
Characteristics of the West African Monsoon (WAM) season, such as its onset and dry spell occurrences, are notoriously difficult to predict. However, these characteristics are key indicators farmers use to decide when to plant crops, having a major influence on their overall yield. While many studies have shown correlations between global sea surface temperatures and characteristics of the WAM season, there are few that effectively implement this information into machine learning (ML) prediction models. This study is focused on predicting dry spells, that is, if there will be a period of consecutive days without rain after the onset of the WAM. We first investigated the best ways to define onset and dry spells and gathered sea surface temperature training data from both real-world observations and a climate simulation model. Then we constructed an adaptive-threshold logistic regression model for dry spell prediction, to which we applied a custom feature selection method and spatial regularization. Using Leave-One-Out cross validation testing, we found significant results in multiple binary classification metrics. These models overcome some limitations that current approaches have, such as being computationally intensive and needing bias correction. We also aim for this study to serve as a framework for ML use in the context of targeted prediction of certain weather phenomena using climatologically relevant variables.
\end{abstract}

\introduction  
The onset of the West African Monsoon (WAM) marks the beginning of intense rainfall that lasts through much of the boreal summer. This seasonal rainfall has a significant impact on the economy since a large portion of economic activity is based on agriculture. In West Africa, the agricultural sector accounts for around 45\% of the workforce and 25\% of gross domestic product \citep{AUC_OECD_2024}. Due to the lack of irrigated farmland, farmers must rely solely on rain for crop growth, making knowledge of rainfall patterns key for a successful growing season. The large majority of yearly rainfall comes during the WAM, so farmers must make important decisions concerning their crops based on when the monsoon season starts, known as the onset date, and any dry spells that might occur thereafter. Although the exact dates for the start of the season may vary across West Africa, there is a coherent large-scale pattern dictated by the shifting of the Inter-Tropical Convergence Zone (ITCZ) northward \citep{sultan2000}. In addition to the relationship between the ITCZ and the WAM, studies have shown teleconnections (large-scale climate relationships across distant regions of Earth) between characteristics of the WAM --- such as early/late onset, dry spells, extreme rainfall, etc. --- and global sea surface temperatures (SSTs) \citep{Salack2013, VariabilityandPredictabilityofWestAfricanDroughts, ImpactsoftheTropicalPacificIndianOceansontheSeasonalCycleoftheWestAfricanMonsoon, OceanicForcingonInterannualVariabilityofSahelHeavyandModerateDailyRainfall, Gbangou2019}. This study aims to evaluate the extent to which global SSTs can inform predictions of the WAM dry spells. There is also the added benefit that these models are not computationally intensive to run, and the predictive signal from SSTs tends to be at quite a large lead time, meaning we can get useful information months before onset. 

Understanding and being able to predict dry spells is crucial for agricultural productivity and economic stability. While weather forecasts have become increasingly available for farmers, many barriers still remain for optimal implementation of this information. These forecasts can be hard to interpret and lacking in relevant predictions \citep{Ingram:2002}. As a consequence, many farmers still tend to judge the rainy season based on traditional knowledge. These indicators have become progressively more untrustworthy though, likely due to climate change \citep{Kalanda-Joshua:2011}. Without reliable prediction methods, farmers are left extremely vulnerable to dry spells, the impacts of which are usually devastating. An extended period of time with little to no rain typically destroys crops, incurring large costs and wasted resources. It also means farmers must labor to replant at some point later in the season, leading to a smaller crop yield and a less productive season. In an attempt to protect farmers against the damaging effects of dry spells, we aim to find a method that can reliably predict them ahead of time.

Current WAM forecasting methods are dominated by numerical weather prediction (NWP) models. These models use a set of initial conditions and equations that govern physical phenomena to forecast weather patterns. These are often made more reliable by running an ensemble of simulations with different initial conditions. Nevertheless, NWP models possess several notable limitations. Firstly, the simulation of these complex interactions requires significant computational resources. For seasonal predictions, not only do computational demands grow, but the error of these models increases with greater lead times as well. Therefore, this can make seasonal predictions infeasible and inaccurate, especially for smaller centers with less access to state of the art models and equipment. Furthermore, West Africa is an area of the world particularly lacking in rainfall data collected on the ground at weather stations. This data is usually the most accurate for rainfall measurements, whereas satellite-derived products almost always have some bias. Sparse station data poses challenges for both the initialization and evaluation of NWP models. Since ground observations are vital for calibrating and validating the precipitation fields produced by these models, the lack of accurate station data degrades the quality of the model's forecast. Lastly, in many studies that investigate accuracy of NWP models for forecasting WAM variables, hindcasts are often used. This means researchers run a simulation from over a period for which there are real-world observations. Since NWP models exhibit bias, a bias-correction method is used based on the observed data in the time period the forecast is run. Yet, issues arise when bias correction is done based on knowledge of weather patterns that would normally not be known at the time of a forecast. In studies where this is done, it is possible that this inflates metrics of skill for the model. This is similar to a ``data leakage" problem in machine learning

The methodology presented in this paper solves some of these issues, while also approaching the problem from a different standpoint. Instead of using large NWP models that predict daily rainfall then calculate downstream events, we make predictions about a specific characteristic of the WAM using only SSTs. We gather data from both real-world and climate simulated sources from oceans across the Earth. Then, we select optimal features for a linear regression model that predicts onsets and a beta regression model that predicts dry spell proportions. After, we implement these predictions into our final logistic regression model to predict where dry spells will occur. Our approach offers a method for making predictions using small datasets with computationally efficient models. In addition to this, our method can make predictions at lead times of up to 6 months. To test our models we use leave-one-out cross-validation (LOOCV), a well-established method for assessing model performance. The idea that using only SSTs could provide us with insights into the behavior of the upcoming monsoon season has major implications. First, it suggests that we can develop a framework for weather prediction in which machine learning methods play an integral role. Additionally, it gives a cheap alternative for targeted predictions that are useful to farmers. With these models being trained specifically for one task, it allows them to be optimized to deliver the best prediction possible. This is in contrast to the indirect prediction seen with NWP models, in which they run simulations of rainfall, then run an onset and dry spell detection algorithm.

\section{Data and General Approach}

\subsection{Area of Interest}
\begin{figure*}[ht!]
    \centering
    \includegraphics[scale=0.55]{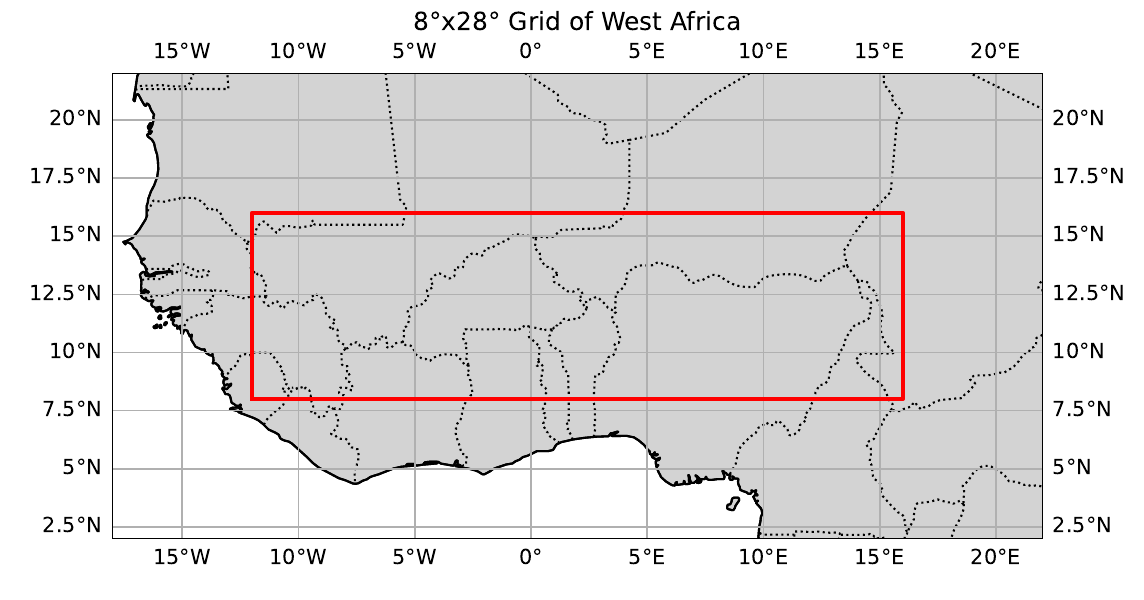}\\
    \caption{Area of Interest (AOI)}
    \label{fig:aoi}
\end{figure*}

We started by defining the area of interest (AOI) for the study, shown in figure \ref{fig:aoi}. We chose a region where the climate is suitable for agriculture and rainfall has significant variability, ensuring that predictions are made for the places that need them most. The area is described by the coordinates, ($8\degree\text{N}-28\degree$N, $12\degree\text{W} -16\degree$E).

\subsection{SST Data}
\begin{figure*}
    \centering
    \includegraphics[scale=0.55]{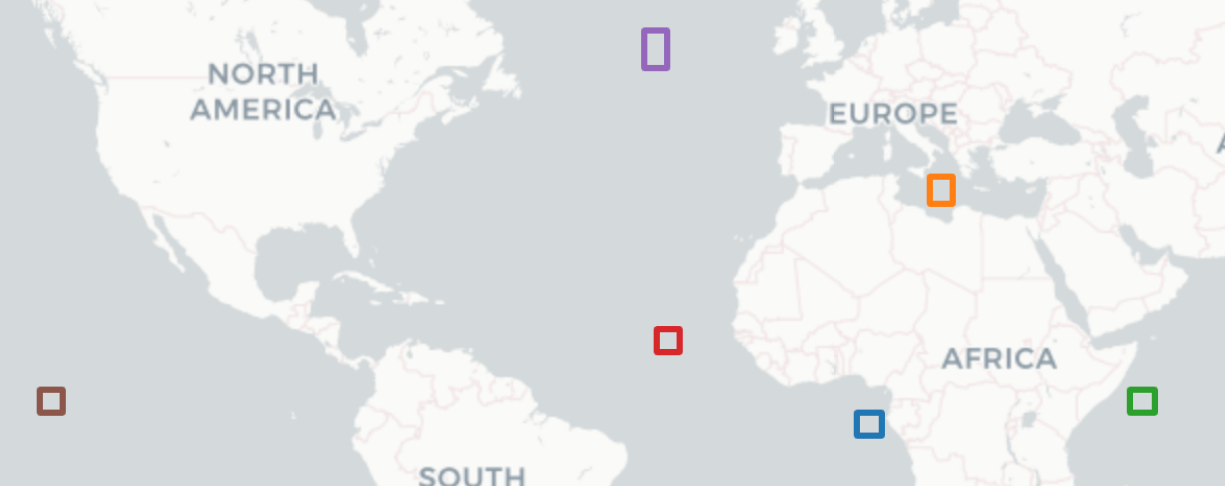}
    \caption{Locations from which we gathered sea surface temperature data}
    \label{fig:fig2}
\end{figure*}

Here we outline the structure of our SST data. At 6 different regions across Earth's oceans, we collected monthly SST averages for the months of a WAM season. These months include September through December for the year before the WAM, and January to March for the year in which WAM happens. We focused on parts of the oceans where temperature anomalies are known to impact weather and climate in West Africa \citep{fonseca:etal:2015,sheen:etal:2017,taschetto2012can,geremew2025unravelling}. The regions considered are: Atlantic ($10^\circ\text{N} - 14^\circ\text{N},\, 30^\circ\text{W} - 26^\circ\text{W}$), North Atlantic ($50^\circ\text{N} - 54^\circ\text{N},\, 32^\circ\text{W} - 28^\circ\text{W}$), Gulf of Guinea ($4^\circ\text{S} - 0^\circ,\, 4^\circ\text{E} - 8^\circ\text{E}$), Indian ($0^\circ - 4^\circ\text{N},\, 50^\circ\text{E} - 54^\circ\text{E}$), Pacific ($0^\circ - 4^\circ\text{N},\, 134^\circ\text{W} - 130^\circ\text{W}$), Mediterranean ($33^\circ\text{N} - 37^\circ\text{N},\, 16^\circ\text{E} - 20^\circ\text{E}$), shown in figure \ref{fig:fig2}. This gives us a data matrix $X_{\text{full}} \in \mathbb{R}^{n \times d}$ where $n$ is the number of samples (available years of data we have) and $d$ is the number of regions multiplied by the number of months we have. Therefore, in total we have $d=42$.

The SST data is obtained from the ECMWF Reanalysis v5 (ERA5) dataset \citep{Hersbach2020} for the years 1981-2024. However, to increase the data size we utilize a climate simulation model data from the Community Earth System Model Version 2 (CESM2), collecting both SST and precipitation data for the years 1935-1980 \citep{cesm2}. CESM2 is an Earth system model that simulates interactions between the atmosphere, ocean, land, and sea ice components to provide historical data of Earth's climate. With these two datasets combined, in total we have $n=90$. The CESM2 data comes at a spatial resolution of $1\degree \times 1\degree$ with a daily temporal resolution. To reduce the dimensionality of the data, we aggregated spatially and temporally. We first selected areas of $4\degree \times 4\degree$ for each sea region, obtaining daily SST values at each of the 16 pixels. Then, at each pixel we calculated the mean temperature over every day of a given month. After, having 16 temperature values per month, we averaged spatially over all the 16 pixels in the grid, leaving us with one temperature value per region per month.

\subsection{Precipitation Data}
For precipitation data we use the Climate Hazards group Infrared Precipitation with Stations (CHIRPS) satellite data \citep{Funk2015CHIRPS}. As before, we blend real-world data (CHIRPS) and climate simulation data (CESM2) for the same years as the SST data. The resolution of the precipitation data is bottle-necked by the resolution of CESM2, which also forces our predictions to have the same resolution. Therefore, we linearly interpolate the CHIRPS data, which originally has spatial resolution $0.25\degree \times 0.25\degree$, to fit the $1\degree \times 1\degree$ resolution of CESM2. This process yields daily precipitation data from 1935-2024 over our AOI, at a $1\degree \times 1\degree$ spatial resolution. This precipitation dataset is then used in the search for onsets and dry spells.

\subsection{Onset and Dry Spell Definitions}
To be able to predict dry spells, we only want to consider them as they happen after the onset, meaning we must take the intermediary step of predicting onsets. Yet, this remains a challenge as the onset of the WAM is known both for its difficulty to predict and for its difficulty in simply defining it. According to \citet{concise} there exist at least 18 distinct definitions in publication. Many of the most agriculturally relevant onset definitions have an explicit criterion that precludes bouts of intense rainfall from being identified as the onset if a dry spell occurs thereafter. In practice though, farmers cannot know beforehand if heavy, persistent rains around the start of the onset will be followed by a dry spell or if the rains will continue consistently for the rest of the season. We take this into account when constructing our definition. In removing the dry spell criterion to search for dry spells themselves, we risk having the definitions identify inconsistent early season rainfall as onsets. Therefore, we aim to find the right balance between onset dates that occur at a time that makes sense to farmers, yet where they are still at a large risk for dry spells after. To that end we take a two-phase approach. First we take care to start our search at a date informed by the local climatology, then we do our onset search using fuzzy logic rules which are not sensitive to strict thresholds. 

The first part comes from \citet{liebmann2012}. Their method involves two main steps. A search start date is defined as one at which the probability of onset occurring earlier is negligibly small. Then, beginning from this date, an algorithm is run to identify the onset from daily rainfall amounts. All of the explicit formulas are included in \citet{Dunning2016}, here we give a brief description on how to find this search start date. We first obtain a long-term average of precipitation per day, given by summing up the total precipitation per year, then dividing by the number of years multiplied by 365, we call this the annual-mean daily average. Next, for each day of the year we sum up all the rainfall on that particular day across all the years in our dataset and then divide by this number of years, giving us the daily average for each day of the year. Then we define cumulative anomaly curve at each point by summing up the differences between the daily average for each day of the year and the annual-mean daily average up until a that point. Then the first day after the minimum of this curve gives us the start of the climatological water season. Finally, to get a date to start our search at, we simply subtract 30 days from this minimum. We used this procedure for each grid cell in our AOI until we had a search start date for each region. 

\begin{figure*}[h]
    \centering
    \includegraphics[scale=0.55]{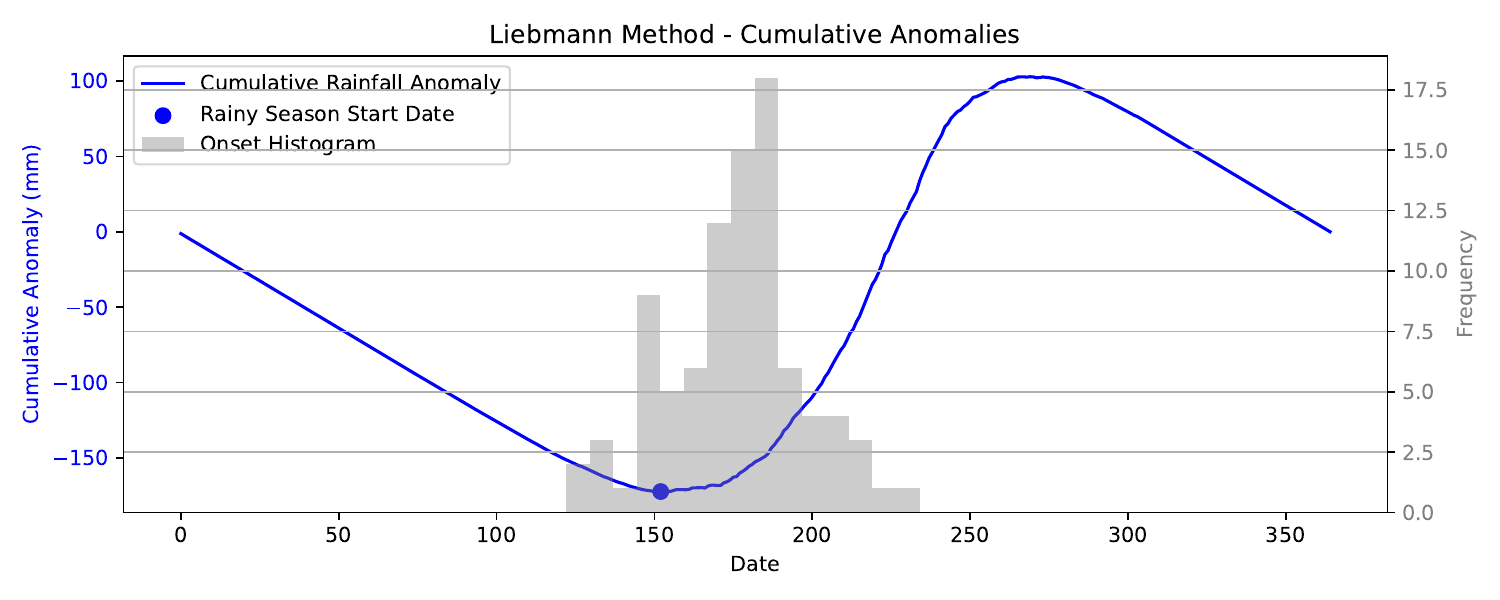}
    \caption{This plot shows the cumulative anomaly curve at pixel 175. The blue dot is the start of the wet season and the histogram shows observed onsets. We would start our search 30 days before the start of the wet season, roughly where the first observed onsets occur.}
    \label{fig:liebmann}
\end{figure*}

For the second part of our onset and dry spell definition, we depart from the Liebmann method and instead switch to a fuzzy logic threshold-based approach. The threshold approach comes from \citet{marteau2009}, and is also used in \citet{Laux2008} where they expanded on it to include fuzzy rules, we do the same here. The standard threshold-based definition for onset is defined as the first day of the year where all of the following conditions are satisfied:
\begin{enumerate}
    \item The cumulative sum of precipitation in five consecutive days is at least $N$ mm
    \item At least $D$ or more of these 5 days must be wet ($\geq$ 1 mm).
\end{enumerate}
Taking after \citet{dodd2001}, we use the above criteria to detect the first realistic potential onset, a period of rainfall that farmers are used to seeing around the beginning of the season. To detect dry spells after this potential onset, we have the last condition:
\begin{enumerate}
    \item[3.] There is a 7-day period of total rainfall less than 5 mm in the succeeding 30 days
\end{enumerate}
coming from \citet{marteau2009}.

The rationale for a fuzzy logic approach is that we have a better chance of picking up the true onset when we do not use strict thresholds. We linearly interpolate between two threshold values for both the variables $D$ and $N$, similar to \citet{Rauch2019}, which gives a score to each value for each day we check. If the product of these two scores is above a certain threshold, that day is designated as the onset. After an onset is detected, criterion (3) is then checked to see if there is a dry spell. A small issue we encounter is, because of varying climate conditions over our AOI, some regions never reach the threshold required to define an onset. The strategy we employ for replacing missing onsets, taken from \citet{boyard-micheau2013regional}, is simply replacing it with the latest onset from that year across the AOI. On average we only needed to replace onsets for 5 pixels out of 224.

\subsection{Exploratory Data Analysis}
We started by exploring the differences between the climate simulated data and CHIRPS data. While many different metrics of accuracy are discussed in \citet{cesm2}, we only calculated the average daily difference for precipitation between the years 1981-2014. We found across our AOI that the average difference between CHIRPS vs climate was $-0.1$ mm, with an absolute difference of about $2.8$ mm. Therefore, we had evidence that simulated data from earlier years would provide us with accurate enough data to capture the relationship between SSTs and the WAM.
\begin{figure*}[h]
    \centering
    \includegraphics[scale=0.45]{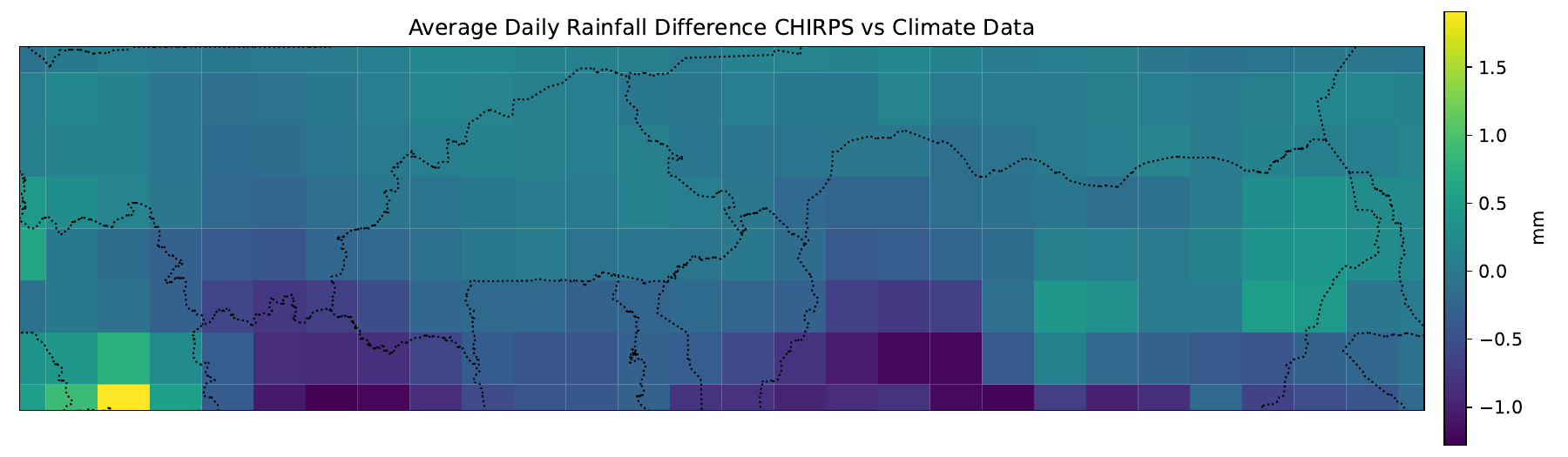}
    \caption{The average daily difference between CHIRPS vs Climate data. 
    }
    \label{fig:climvschirs}
\end{figure*}

\begin{figure*}[h]
    \centering
    \includegraphics[scale=0.45]{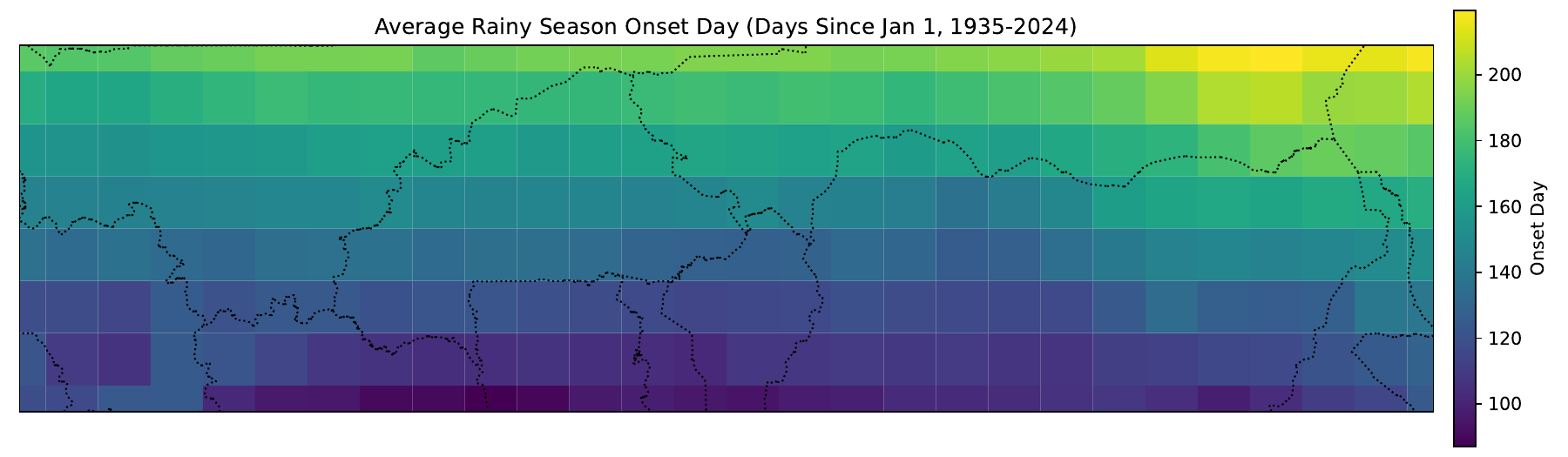}\\
    \includegraphics[scale=0.45]{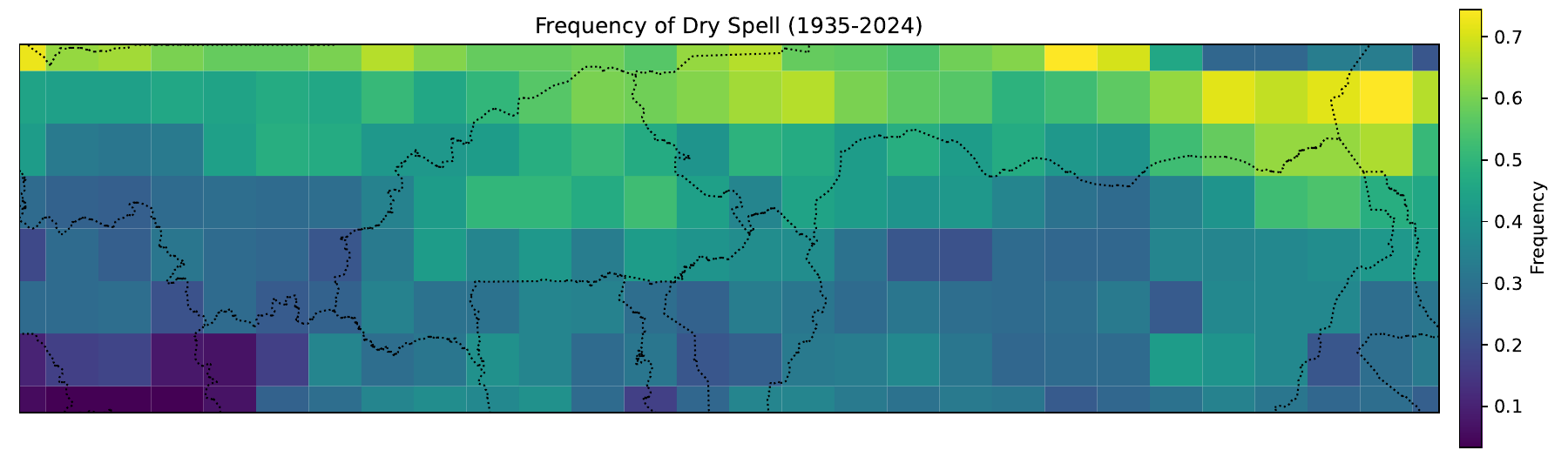}
    
    \caption{Average onset day (top). The observed frequency of dry spells (bottom) are given as well.}
    \label{fig:avg}
\end{figure*}
To characterize regional WAM onsets and dry spells, we analyzed the outputs of our onset and dry spell detection algorithm. To start, we simply found the mean and standard deviation of onsets for each grid cell over the 90 years we had available. The pattern of onsets follows what we expected to see. As shown in figure \ref{fig:avg}, the onsets of the rainy season have a northward gradient consistent with the movement of the ITCZ. We also see that dry spells vary significantly over the region but with an increased frequency of dry spells further North where the climate becomes more arid. Averaging over all years of data and all grid points shows any region has a 39\% chance of dry spell. The average standard deviation of a region's dry spell probability/frequency is 46\%. These numbers drop to 24\% and 40\% respectively when taking into account only real world data.

\subsection{Feature Building}
\begin{figure*}[h!]
    \centering
    \includegraphics[scale=0.45]{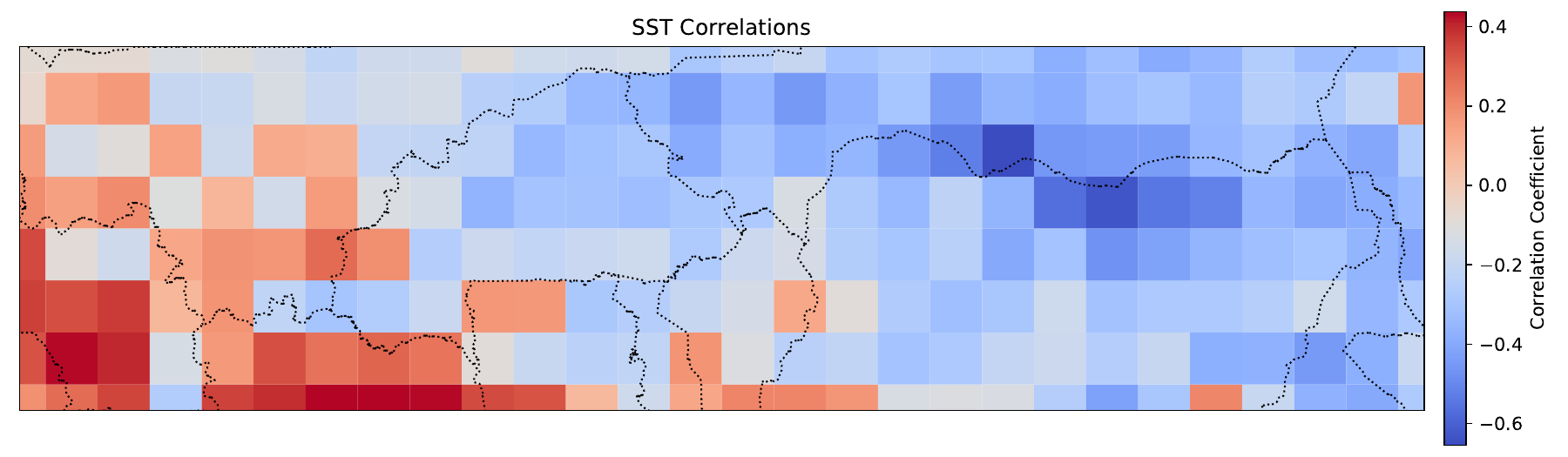}\\
    \includegraphics[scale=0.45]{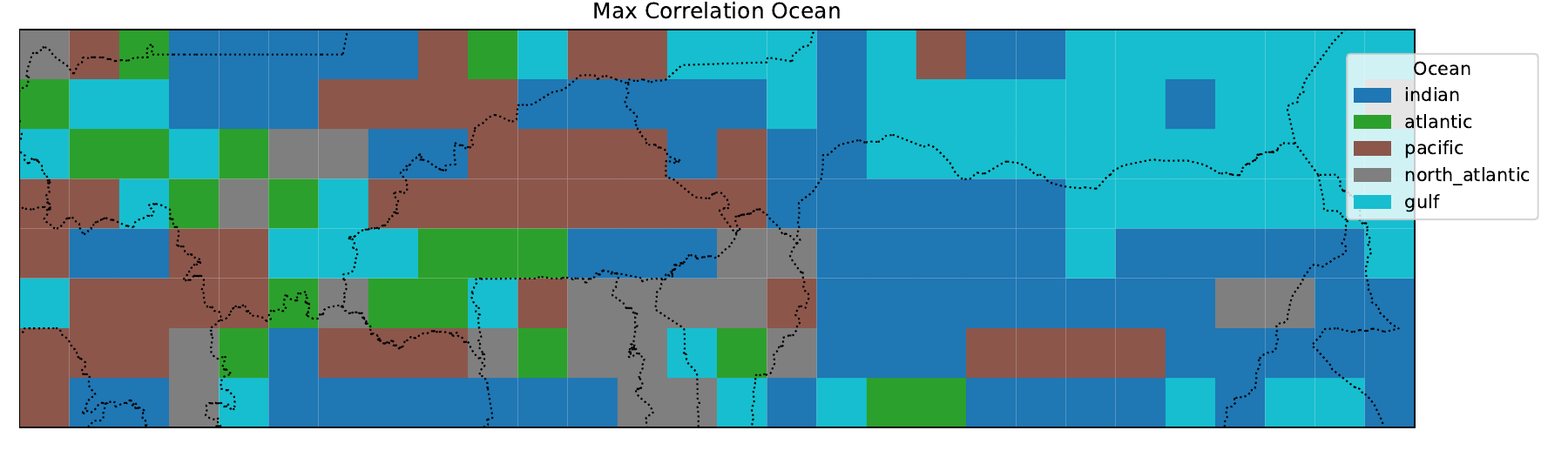}
    
    \caption{The maximum correlation coefficient for onsets (top), and the ocean areas they come from (bottom).}
    \label{fig:sst}
\end{figure*}

One of our key contributions is the way we selected features for our models. To determine what features to use we explored the correlations between SSTs and characteristics of the WAM. We began by running a Pearson correlation test for each region between its onset dates and each column of our SST data matrix $X_\text{full}$, giving $d$ correlations coefficients for every region. Again, being aware of the fact that the simulated climate data might skew results, we ran these tests on both the full dataset and the subset of it that contains strictly real-world data. We report the findings only from tests on the real-world data since these are the tests we use to determine the features of our model. Our results show we get up to a $-0.59$ correlation with an average absolute correlation over all regions of $0.34$. Figure \ref{fig:sst} reveals where the strongest correlations are in the AOI, and which ocean areas the strongest correlations come from. We observe fairly strong signal from certain areas across the AOI. Specifically, we see a clear pattern of negative correlations with the Indian Ocean and the Gulf of Guinea on the Eastern side, however no discernible pattern further to the West. We also ran correlation tests between our collected SST data and the proportion of our AOI that had a dry spell that year. From these tests we see up to a $-0.54$ correlation, indicating that there exists some non-trivial relationship between SSTs and the amount of dry spells that occur in a given year across West Africa. The two areas that gave the best signal were the Gulf of Guinea in October and the Mediterranean in September. This information came in very useful when adapting our threshold for dry spell predictions, as discussed in section 3. 

Based on the results of these correlation tests we select features for the onset and dry spell proportion models. First, we choose one common month to select our features from, looking as early as possible. Then, for each pixel we use the two regions that have the highest correlation with the onset dates in that region. For the onset model we obtain a data matrix for each pixel $i$, $X^{(i)}_{\text{ons}} \in \mathbb{R}^{90\times2}$. We later add a constant column of 1's to the beginning of this matrix to fit the intercept in our model. For the dry spell proportion dataset we select the oceans with the highest correlation, giving us $X_{\text{prop}} \in \mathbb{R}^{90\times2}$.

\section{Models and Prediction Method}
This section details the models we created to make predictions for the onset dates and dry spells. As input to our model we have the training data $X_{\text{full}} \in \mathbb{R}^{n \times d}$, and our target variables. For ease of computation we flatten the pixels of the $8\degree\times28\degree$ AOI into a vector of length $224$. Our first response variable is $y_{\text{ons}}$. It is in the form of days after January 1st. Although initially these are discrete time steps, through standardizing it (subtracting by the mean and dividing by the standard deviation per pixel) we can treat it as a continuous variable, and thus do linear regression on it. We also have dry spell, $y_{\text{ds}}$, which is binary-valued, either 0 for no dry spell or 1 for dry spell. For both we have observations for each pixel in the grid and we have 90 years worth of data giving  
\[
y_{\text{ons}} \in \mathbb{R}^{90\times224}, \quad  y_{\text{ds}} \in \{0, 1\}^{90 \times 224}.
\]
We also add one more step to the prediction pipeline by predicting the proportion of pixels that will have a dry spell for a given year. This is a useful step for adjusting the final dry spell model. For the proportion of dry spells we have data 
\[
y_{\text{prop}} \in [0,1]^{90}.
\]

\subsection{Onset Model}
As mentioned in section 2.6, we have a custom feature building process for each pixel which we will now employ. To use this data to make predictions we start by assuming that our target variable follows
\[
y_{\text{ons},t}^{(i)}|X_{\text{ons}, t}^{(i)} \sim \mathcal{N}(X_{\text{ons}, t}^{(i)} \beta^{(i)}, \, \sigma^2) \quad \text{for } i=1,\dots, 224,
\]
where $y_{\text{ons}, t}^{(i)}$ is the true onset at pixel $i$ for year $t$, $X_{\text{ons}, t}^{(i)}$ is the row vector obtained from the $t$th row of $X_{\text{ons}}^{(i)}$ with a 1 in the first entry for the intercept, $\beta^{(i)} \in \mathbb{R}^{3}$ is a column vector of regression coefficients, and $\sigma^2$ is some variance we assume remains constant. This leaves us with the task of learning the optimal coefficients with the best fit (i.e. minimize loss).

Using our EDA and what we know about the way the ITCZ moves, it obvious that onsets exhibit a spatial relationship. If two regions neighbor each other they will both have precipitation events that lead to onsets around the same time. Based on this fact we implement a slightly modified version of what is known as total variation (TV) regularization. First we build an adjacency list: for each pixel, we get the indices of the pixels above, below, and on either side of it. Then based on this list we say $i \sim j$ if region $i$ is neighbors with region $j$, thus the notation under the summation in the regularization term indicates that we sum over all pairs of neighbors. Normally, we would apply this regularization to the parameters (regression coefficients in this case) of our model. In this case though, since we use custom data for each pixel, the coefficients do not correspond to the same features and we instead regularize the predictions. So, we fit all the $P$ models jointly by minimizing the regularized loss function 
\[
\mathcal{L}_{\text{ons}}\left(\beta^{(1)}, \dots, \beta^{(P)}\right) = \frac{1}{P}\sum_{i=1}^{P} \left[\frac{1}{B}\sum_{t=1}^B\left( X_{\text{ons}, t}^{(i)} {\beta}^{(i)} - y_{\text{ons},t}^{(i)} \right)^2\right] + \lambda_{\text{ons}} \, \sum_{i \sim j} \sum_{t=1}^B \left| X_{\text{ons}, t}^{(i)} {\beta}^{(i)} - X_{\text{ons}, t}^{(j)} {\beta}^{(j)} \right|,
\]
where $P$ is the number of grid cells in our AOI and $B$ is the batch size used when training. We trained this model using backpropagation and an Adam optimizer to find the coefficients that minimize our loss function. The hyperparameters for the model are all fine-tuned using a Bayesian hyperparameter search. Our model's performance is then evaluated using leave-one-out cross-validation and we report the outcomes of our tests in the Results section. 

\subsection{Dry Spell Proportion}
This model is created to predict the proportion of dry spell across the area (i.e. the percentage of grid cells in which a dry spell occurs) for a given year. To do this we subset another part of our original dataset, $X_{\text{full}}$, taking the Gulf of Guinea SSTs for the month of October and the Mediterranean SSTs for the month of September. As mentioned in the EDA section, for both of these region/month combinations we found a -0.54 correlation between their SSTs and dry spell proportion. The interpretation being, the colder these two seas were, the higher the proportion of dry spell across West Africa. With this data we fit a beta regression model to ensure our predictions will be bounded between 0 and 1. In this model we make the assumptions that
\[
y_{\text{prop}, t} \mid X_{\text{prop}, t} \sim \text{Beta}(\mu_t\phi, (1-\mu_t)\phi), \quad \text{with the link function,} \quad
\text{logit}(\mu_t) = X_{\text{prop}, t}\beta_{\text{prop}}
\]
where $y_{\text{prop}, t}$ is the dry spell proportion for time $t$, $X_{\text{prop}} \in \mathbb{R}^{90 \times 3}$ is the subsetted SST data we described above starting with a column of 1's for the intercept and $X_{\text{prop}, t}$ is the $t$th row vector, $\beta_{\text{prop}} \in \mathbb{R}^3$ are the regression coefficients including the intercept, $\mu_t$ is the mean of the distribution for year $t$ that is linked to the linear predictor, and $\phi$ is a precision parameter assumed to be constant. This $y_{\text{prop}, t}$ will be used as a key part of our logistic regression model. 

\subsection{Dry Spell Model}
The dry spell model is aimed at predicting whether dry spells occur at each region for a given year. To accomplish this we use a logistic regression model and again add TV regularization. As with the other models, we subsetted our SST data matrix $X_{\text{full}}$, choosing SSTs from the month of October because they provided the strongest signal at a significant lead time. Also, we reduced the amount of features to just the regions: Indian, Gulf of Guinea, Mediterranean, and North Atlantic. The other features include two constant columns of the pixel's latitude and longitude, and the onset date. Note that while the onset model dataset varied across pixels, this one remains constant across pixels, except for the last two columns. The true onset is used in training, while our predicted onset is used during testing, ensuring no data leakage. 

The modeling assumption we make is, 
\[
y^{(i)}_{\text{ds},t}|X_{\text{ds},t}^{(i)} \sim \text{Bernoulli}\left(\frac{e^{X_{\text{ds},t}^{(i)}\theta^{(i)} + b^{(i)}}}{1 + e^{ X_{\text{ds},t}^{(i)}\theta^{(i)} + b^{(i)}}}\right), 
\]
where $y^{(i)}_{\text{ds},t} \in \left\{0, 1 \right\}$ is whether a dry spell occurs or not at region $i$ for year $t$,  $X_{\text{ds}}^{(i)} \in \mathbb{R}^{90 \times 6}$ is the data we subsetted from our original SST data to use as features for our model so $X_{\text{ds},t}^{(i)} \in \mathbb{R}^{6}$ is the row vector of data at year $t$ and pixel $i$, and $\theta^{(i)} \in \mathbb{R}^6$ is the vector of coefficients for pixel $i$ with the intercept $b^{(i)}$ as well. When this is done in practice, we get the probabilities for our predictions by calculating $\sigma(z_t^{(i)})$, where $\sigma(z_t^{(i)}) = 1/1+e^{-z_t^{(i)}}$ is the sigmoid function and $z_t^{(i)} = X_{\text{ds},t}^{(i)}\theta^{(i)}$. Then we adaptively find a threshold $T \in [0,1]$ based on the dry spell proportion prediction. Using $T$ and the probability we calculated, we decide whether to classify the prediction as a dry spell or not. 

The regularization we put on the dry spell model is slightly different than that of the onset model. This time we make predictions for each pixel based on the same data, so we can directly regularize the coefficients of two different pixels. Therefore, the expression for our TV regularization is 
\[
\text{TV}_{\text{ds}}(\theta^{(1)}, \dots, \theta^{(P)}) = \sum_{i \sim j} \sum_{k=1}^d  \left( \theta^{(i)}_k - \theta^{(j)}_k \right)^2,
\]
where $\theta^{(i)}_k \in \mathbb{R}$ is the $k$th entry in the parameter vector for the $i$th pixel's logistic regression model. This expression is more consistent with standard TV regularization, although instead of using the absolute value we square the difference. For training we use the binary cross entropy (BCE) loss, which is the same as the negative log likelihood of the logistic regression model. The BCE loss combined with the TV regularization gives us
\[
\mathcal{L}\left(\theta^{(1)}, \dots, \theta^{(P)}, \; b^{(1)},  \dots, b^{(P)}\right) = \frac{1}{P}\sum_{i=1}^{P} \left[- \frac{1}{B}\sum_{t=1}^B y_{\text{ds},t}^{(i)} \log \sigma(z_t^{(i)}) + (1 - y_{\text{ds},t}^{(i)}) \log (1 - \sigma(z_t^{(i)}))\right] +  \lambda_{\text{ds}}\, \text{TV}_{\text{ds}}(\theta^{(1)}, \dots, \theta^{(P)}).
\]
From here we obtain the learned parameters $\theta^{(1)}, \dots, \theta^{(P)}, \; b^{(1)},  \dots, b^{(P)}$, which we use to make predictions.

\subsection{Prediction}
After training, we want to use the optimized model parameters to make dry spell predictions. When we make a prediction for a given year, we assume all the response variables are as yet unknown, and we only have access to that year's SSTs. Having obtained the learned regression coefficients $\hat{\beta}^{(1)}, \dots, \hat{\beta}^{(P)}$, the prediction of the onset for year $t^\prime$ at pixel $i$ is given by 
\[
\hat{y}^{(i)}_{\text{ons},t^\prime} = X_{\text{ons}, t^\prime}^{(i)} \hat{\beta}^{(i)}.
\]
These predictions are later used as features for the dry spell predictions. 

Next we predict the dry spell proportion. After we fit the beta regression model we have the estimated parameters $\hat{\beta}_{\text{prop}}$. We use this and the new sample to compute the linear predictor, then we plug it in to the inverse link function, i.e. the sigmoid function, shown as 
\[
\hat{y}_{\text{prop},t^\prime} = \sigma \left(X_{\text{prop}, t^\prime} \hat{\beta}_{\text{prop}}\right).
\]

Finally we make our prediction for the dry spell using the predictions from previous steps. The onset predictions are used during training while the dry spell proportion prediction is used at test time. First notice that our model's output is $\sigma(z_{t^\prime}^{(i)})$, the probability a of dry spell, which we must dichotomize. Traditionally, if $\sigma(z_{t^\prime}^{(i)}) \geq 0.5$ then the prediction is 1, otherwise it is 0. However, likely due to our small sample size, the logistic regression model at the $0.5$ threshold is not well-calibrated. To solve this issue we use an adaptive threshold. We start by interpreting the values of $\sigma(z_{t^\prime}^{(i)})$ for all pixels $i = 1, \dots, 224$ as an empirical distribution of predicted probabilities for year $t^\prime$. There exists a CDF $F$ for this distribution where $F(T) = 1 - {y}_{\text{prop}, t^\prime}$, meaning $1 - {y}_{\text{prop}, t^\prime} \cdot 100\%$ of probabilities are less than $T^\ast$ and ${y}_{\text{prop}, t^\prime} \cdot 100\%$ are greater than $T^\ast$. If we knew ${y}_{\text{prop}, t^\prime}$ then we could use the inverse CDF, also known as the quantile function $Q$, to get $Q(1-{y}_{\text{prop}, t^\prime}) = T^\ast$, which means we could perfectly predict how many dry spells would occur that year. Since this is impossible, we settle for using our prediction to find $Q(1-\hat{y}_{\text{prop}, t^\prime}) = T$. Notice $T$ will likely not be equal to $T^\ast$, however if our estimate of the proportion is good, then so will our estimate of $T^\ast$. This now means $(1-\hat{y}_{\text{prop}, t^\prime}) \cdot 100 \%$ of predicted probabilities are less than $T$, and the other $\hat{y}_{\text{prop}, t^\prime} \cdot 100 \%$ of probabilities above $T$. Therefore we assign, 
\[
 \hat{y}^{(i)}_{\text{ds},t^\prime} = \begin{cases}
0 & \text{if } \sigma(z_{t^\prime}^{(i)}) < T \\
1 & \text{if } \sigma(z_{t^\prime}^{(i)}) \geq T.
\end{cases}
\]
This ensures we predict a proportion of dry spells across the AOI that is close to the actual proportion, and the regions with the highest probability of dry spells will be predicted to have a dry spell. Empirically, this strategy, as is shown in the Results section, gives the best balance between precision and recall metrics.

\section{Results}
Here we present results obtained from our dry spell model, with a note on the performance of our onset and dry spell proportion model as well. To obtain these results we used Leave-One-Out Cross-Validation (LOOCV). This approach is standard when working with very small sample sizes, since train–validation–test splits and even $k$-fold cross-validation require holding out too much data for testing, which can limit the information available for training. One iteration of LOOCV consists of leaving one year out of our training data, then training on the full 89 years left over and making a prediction for the left out year. This is repeated for each of the 44 years for which we have real-world data, so our reference to compare our prediction to has a real-world value. Each iteration produces a different set of results which are then aggregated, and the final results are reported below.

\subsection{Onset Results}
As mentioned, we did LOOCV to test our onset model. Meaning we created onset predictions for all grid points on a given year based on the trained model and the SST data from September of that given season, then we repeated this for 44 years. We compared each prediction to the ground truth value to measure the accuracy of our model. One common accuracy metric used for regression tasks is mean absolute error (MAE), given by the formula
\[
\text{MAE} = \frac{1}{n_r \cdot P}\sum_{t=1}^{n_r} \sum_{i=1}^{P} \left| y_{\text{ons},t}^{(i)} - \hat{y}^{(i)}_{\text{ons},t} \right|,
\]
where $n_r=44$ is the number of years for which we have real-world data. Using this methodology the lowest MAE we obtained was 11.5 days. This means that on average, the absolute difference between the predicted value and the true value is 11.5. However, this number is misleading since one can simply predict the mean onset date at a give grid cell and obtain an MAE of 11.8. So, to further evaluate this model for its skill we used several evaluation metrics, many of the same used in \citet{Rauch2019}. Although we found that in many cases the model predicts values close to the mean, there were pixels with greater accuracy and significant skill. In fact, in comparison to \citet{Rauch2019}, almost all of these metrics --- including spatial bias, root mean squared error (RMSE), mean spatial correlation, mean spatial anomaly correlation, yearly anomaly correlation, correlation between predictions and target, skill vs. climatology --- were very similar and in many cases better. So, despite our model not having enough predictive power to give meaningful predictions to farmers, the aforementioned metrics indicated some non-insignificant skill, and therefore we kept these predictions and used them as a feature in the dry spell model. An interesting avenue for future research would be to add more informative features on top of SSTs to see how accurate these predictions can be while still making them at a large lead time. 

\subsection{Dry Spell Proportion Results}
For the dry spell proportion model we also used the MAE to evaluate the predictions. We found an MAE of about $6.7\%$ for our model whereas predicting the mean each year yields an MAE of about $17\%$. This among other metrics we evaluated show that our model is a good fit and has significant predictive power. This contributes to the ability of our model to balance the number of predictions in each category.

\subsection{Dry Spell Results}
We take a very similar approach as the onset model to getting test results. We again use LOOCV, testing on all years we have real world data for. For evaluation of our model, the most important metrics come in the form of a classification report typical for binary classification problems. We care about the metrics: precision, recall, and $F_1$-score. The formulas for these are as follows,
\begin{align*}
\text{Precision} &= \frac{T\!P}{T\!P + F\!P}, \quad
\text{Recall} = \frac{T\!P}{T\!P + F\!N}, \quad F_1 = \frac{2 \cdot \text{Precision} \cdot \text{Recall}}{\text{Precision} + \text{Recall}},
\end{align*}
where $T\!P$ = True Positives (positives we predict that are correct), $F\!P$ = False Positives (positives we predict whose ground truth value is 0) and $F\!N$ = False Negatives (zeros we predict whose true value is 1). Thus, precision represents the proportion of predicted positives that are correct, while recall measures the proportion of actual positives that were correctly identified. $F_1$ score is the harmonic mean of the two. The full report is given in Table 1, where the macro row is an average of both classes' scores, the weighted row is a weighted average of both classes' scores based on the class size, and support gives the class counts.
\begin{table*}[h!]
\centering
\scalebox{1.25}{
\begin{tabular}{lcccc}
\hline
Class & Precision & Recall & F1-Score & Support \\
\hline
0 & 0.83 & 0.80 & 0.81 & 7479 \\
1 & 0.44 & 0.50 & 0.46 & 2377 \\
\hline
Accuracy &  &  & 0.72 & 9856 \\
Macro Avg & 0.63 & 0.65 & 0.64 & 9856 \\
Weighted Avg & 0.74 & 0.72 & 0.73 & 9856 \\
\hline
\end{tabular}
}
\caption{Classification report for dry spell model tests}
\end{table*}

These results are difficult to compare directly with existing studies since, to our knowledge, we take a novel approach to the problem. While previous work has examined various characteristics or drivers of dry spells, we are not aware of any studies that predict a binary outcome indicating the occurrence or absence of a dry spell. However, we can compare this against a baseline of just predicting 0's since that is the majority category. Recall we have an average of 0.24 dry spells across the AOI over 44 years of observed data. Naturally this yields an overall accuracy of 76\%, only a small improvement to our model's accuracy of 72\%. Despite this, raw accuracy can be misleading. For insight into our model's performance we look at the class 1 row of the classification report table. This tells us that our precision is 44\%, recall is 50\% and $F_1$ score is 46\%. This is interpreted as meaning that out of all the dry spells our model predicted, it was corect 44\% of the time. Additionally, out of all the dry spells that occurred over all our test years, the model predicted the dry spell 50\% of the time. We also calculate the Receiver Operating Characteristic – Area Under the Curve (ROC-AUC), which provides an aggregate measure of performance across all possible classification thresholds. Our model achieves an ROC-AUC of 0.65, indicating a moderate ability to distinguish between years with and without dry spells. This value again shows that our model is better than predicting only 0's (which would yield 0.5). These are strong results considering the small sample size and small number of explanatory variables, and it confirms what studies like \citep{Salack2013} and others have suggested, that SSTs influence WAM characteristics like dry spell.

\conclusions 
This study's focus was to predict dry spell occurrences after the onset of the WAM and to evaluate the effectiveness of global sea surface temperature (SST) teleconnections as features. We use linear, beta, and logistic regression models with custom spatial regularization and a feature selection technique based off correlations. Fitting these models allowed us to make future predictions about where dry spells would occur for every $1\degree \times 1 \degree$ cell in an $8\degree \times 12 \degree$ grid. These predictions are made based off SSTs that are observed no later than October the year before the monsoon season. To gather training data we implemented a hybrid method of known onset definitions giving both potential onsets and dry spells. We also used data from climate simulated models to add more samples to our training set, thereby enhancing the model's performance. To test our models, we utilized LOOCV for every year of real-world data we had. Our tests showed significant results across a number of binary classification metrics. This work represents an initial step toward developing a framework for models capable of predicting characteristics of the WAM months in advance. Such a framework could help farmers prepare more effectively for the coming season and enable weather prediction models to refine their forecasts as the season approaches, improving overall accuracy. Ultimately, the goal of this study is to demonstrate that large-scale sea surface temperature (SST) patterns contain valuable information that can be leveraged in ML models for predicting the behavior the WAM.

Despite our promising results, there are several limitations to the approach we took and the models that were developed. Firstly, an inherent limitation is the lack of data. It is not the lack of features, but rather the lack of samples ---  which in our case are the years of data we have --- that is the main issue. While there are statistical methods for dealing with this problem, it will always be a limiting factor in the ML model's ability to make accurate predictions. Without sufficient samples it is difficult for the model to learn the correct relationship between the features and the target variables, no matter what model is chosen. Additionally, we are limited with regards to the spatial resolution at which we can make predictions. As mentioned, the resolution of our climate simulated data is $1\degree \times 1\degree$, meaning we had to regrid our CHIRPS data to match this and ultimately, our predictions come at the same scale. Within a region of that size, there still exists variability in weather patterns and thus the prediction might not be accurate for the all parts of the region. One potential solution is, if it exists, to use data with a finer resolution. However, this will scale up the size of the model multipicatively, requiring more computational resources. Lastly, although we explored several methods for uncertainty quantification --- that is, assigning confidence levels to predictions ---  we were unable to obtain any informative results, again likely due to the small sample size.  

Based off these results we believe there are many avenues for future research that seem promising. The natural progression from the frequentist model we constructed would be to take a Bayesian approach. For one, it would give a natural way of doing uncertainty quantification that would be much more informative to farmers than a single prediction without measure of how sure we are of it. Also, Bayesian models are known to more robust to small sample sizes, incorporating prior information. Another direction for future research would be to integrate these predictions with those from NWP models. While neither approach is perfect, combining them could provide a useful measure of uncertainty: when both models agree, the forecast is likely to be more reliable, whereas large discrepancies between them could signal lower confidence in the prediction. Lastly, getting accurate predictions for onset would also be useful in tadem with the dry spell predictions, it seems as though adding more features to the model would be necessary though, and more complex modeling approaches can be taken that still fit within an ML framework.

\begin{acknowledgements}
We gratefully acknowledge support from Boston University's Undergraduate Research Opportunities Program and Boston University's Newbury Center. 
\end{acknowledgements}

\bibliographystyle{copernicus}
\bibliography{references}

\end{document}